\documentclass[conference,10pt]{IEEEtran}
\usepackage{url}
\usepackage[utf8]{inputenc}
\usepackage{xcolor}
\usepackage{amsmath}
\usepackage{amssymb}

\usepackage[acronyms,nonumberlist,nopostdot,nomain,nogroupskip]{glossaries}
\usepackage{tablefootnote}
\usepackage{booktabs}

\usepackage{tabularx}

\usepackage{tikz}
\usepackage{pgfplots}
\pgfplotsset{compat=newest} 
\pgfplotsset{plot coordinates/math parser=false} 
\newlength\fheight
\newlength\fwidth
\usetikzlibrary{plotmarks,patterns,decorations.pathreplacing,backgrounds,calc,arrows,arrows.meta,spy,matrix}
\usepgfplotslibrary{patchplots,groupplots}
\usepackage{tikzscale}
\usepackage{hyperref}

\newif\ifexttikz
\exttikzfalse
\ifexttikz
	\usetikzlibrary{external}
\fi

\usepackage{multirow}
\usepackage{tkz-kiviat}

\usepackage[font=scriptsize]{subcaption}
\usepackage[font=footnotesize]{caption}

\usepackage{mathtools}

\newacronym{3gpp}{3GPP}{3rd Generation Partnership Project}
\newacronym{adc}{ADC}{Analog to Digital Converter}
\newacronym{5g}{5G}{5th generation}
\newacronym{aimd}{AIMD}{Additive Increase Multiplicative Decrease}
\newacronym{am}{AM}{Acknowledged Mode}
\newacronym{amc}{AMC}{Adaptive Modulation and Coding}
\newacronym{aqm}{AQM}{Active Queue Management}
\newacronym{awgn}{AGWN}{Additive White Gaussian Noise}
\newacronym{balia}{BALIA}{Balanced Link Adaptation}
\newacronym{bdp}{BDP}{Bandwidth-Delay Product}
\newacronym{bf}{BF}{Beamforming}
\newacronym{cc}{CC}{Congestion Control}
\newacronym{cdf}{CDF}{Cumulative Distribution Function}
\newacronym{cn}{CN}{Core Network}
\newacronym{cqi}{CQI}{Channel Quality Information}
\newacronym{cp}{CP}{Control Plane}
\newacronym{csirs}{CSI-RS}{Channel State Information - Reference Signal}
\newacronym{dc}{DC}{Dual Connectivity}
\newacronym{dce}{DCE}{Direct Code Execution}
\newacronym{dci}{DCI}{Downlink Control Information}
\newacronym{dl}{DL}{Downlink}
\newacronym{dmr}{DMR}{Deadline Miss Ratio}
\newacronym{dmrs}{DMRS}{DeModulation Reference Signal}
\newacronym{e2e}{E2E}{End-to-End}
\newacronym{ecn}{ECN}{Explicit Congestion Notification}
\newacronym{edf}{EDF}{Earliest Deadline First}
\newacronym{enb}{eNB}{evolved Node Base}
\newacronym{epc}{EPC}{Evolved Packet Core}
\newacronym{es}{ES}{Edge Server}
\newacronym{fdma}{FDMA}{Frequency Division Multiple Access}
\newacronym{fdd}{FDD}{Frequency Division Duplexing}
\newacronym[firstplural=Radio Access Technologies (RATs)]{rat}{RAT}{Radio Access Technology}
\newacronym{fs}{FS}{Fast Switching}
\newacronym{ftp}{FTP}{File Transfer Protocol}
\newacronym{gnb}{gNB}{Next Generation Node Base}
\newacronym{harq}{HARQ}{Hybrid Automatic Repeat reQuest}
\newacronym{hetnet}{HetNet}{Heterogeneous Network}
\newacronym{hh}{HH}{Hard Handover}
\newacronym{hol}{HOL}{Head-of-Line}
\newacronym{ia}{IA}{Initial Access}
\newacronym{imt}{IMT}{International Mobile Telecommunication}
\newacronym{iot}{IoT}{Internet of Things}
\newacronym{los}{LOS}{Line of Sight}
\newacronym{lte}{LTE}{Long Term Evolution}
\newacronym{m2m}{M2M}{Machine to Machine}
\newacronym{mac}{MAC}{Medium Access Control}
\newacronym{mc}{MC}{Multi-Connectivity}
\newacronym{mcs}{MCS}{Modulation and Coding Scheme}
\newacronym{mec}{MEC}{Mobile Edge Cloud}
\newacronym{mi}{MI}{Mutual Information}
\newacronym{mimo}{MIMO}{Multiple Input, Multiple Output}
\newacronym{mmwave}{mmWave}{millimeter wave}
\newacronym{mptcp}{MPTCP}{Multipath TCP}
\newacronym{mr}{MR}{Maximum Rate}
\newacronym{mss}{MSS}{Maximum Segment Size}
\newacronym{mtd}{MTD}{Machine-Type Device}
\newacronym{mtu}{MTU}{Maximum Transmission Unit}
\newacronym{nfv}{NFV}{Network Function Virtualization}
\newacronym{nlos}{NLOS}{Non Line of Sight}
\newacronym{nr}{NR}{New Radio}
\newacronym{ofdm}{OFDM}{Orthogonal Frequency Division Multiplexing}
\newacronym{pdcch}{PDCCH}{Physical Downlonk Control Channel}
\newacronym{pdcp}{PDCP}{Packet Data Convergence Protocol}
\newacronym{pdsch}{PDSCH}{Physical Downlink Shared Channel}
\newacronym{pdu}{PDU}{Packet Data Unit}
\newacronym{pf}{PF}{Proportional Fair}
\newacronym{pgw}{PGW}{Packet Gateway}
\newacronym{phy}{PHY}{Physical}
\newacronym{pbch}{PBCH}{Physical Broadcast Channel}
\newacronym[plural=\gls{mme}s,firstplural=Mobility Management Entities (MMEs)]{mme}{MME}{Mobility Management Entity}
\newacronym{prb}{PRB}{Physical Resource Block}
\newacronym{pss}{PSS}{Primary Synchronization Signal}
\newacronym{pucch}{PUCCH}{Physical Uplink Control Channel}
\newacronym{pusch}{PUSCH}{Physical Uplink Shared Channel}
\newacronym{rach}{RACH}{Random Access Channel}
\newacronym{ran}{RAN}{Radio Access Network}
\newacronym{red}{RED}{Random Early Detection}
\newacronym{rf}{RF}{Radio Frequency}
\newacronym{rlc}{RLC}{Radio Link Control}
\newacronym{rlf}{RLF}{Radio Link Failure}
\newacronym{rrc}{RRC}{Radio Resource Control}
\newacronym{rrm}{RRM}{Radio Resource Management}
\newacronym{rr}{RR}{Round Robin}
\newacronym{rs}{RS}{Remote Server}
\newacronym{rsrp}{RSRP}{Reference Signal Received Power}
\newacronym{rss}{RSS}{Received Signal Strength}
\newacronym{rtt}{RTT}{Round Trip Time}
\newacronym{rw}{RW}{Receive Window}
\newacronym{rx}{RX}{Receiver}
\newacronym{sa}{SA}{standalone}
\newacronym{sack}{SACK}{Selective Acknowledgment}
\newacronym{sap}{SAP}{Service Access Point}
\newacronym{sch}{SCH}{Secondary Cell Handover}
\newacronym{scoot}{SCOOT}{Split Cycle Offset Optimization Technique}
\newacronym{sdma}{SDMA}{Spatial Division Multiple Access}
\newacronym{sinr}{SINR}{Signal to Interference plus Noise Ratio}
\newacronym{sm}{SM}{Saturation Mode}
\newacronym{snr}{SNR}{Signal to Noise Ratio}
\newacronym{son}{SON}{Self-Organizing Network}
\newacronym{ss}{SS}{Synchronization Signal}
\newacronym{srs}{SRS}{Sounding Reference Signal}
\newacronym{sss}{SSS}{Secondary Synchronization Signal}
\newacronym{tb}{TB}{Transport Block}
\newacronym{tcp}{TCP}{Transmission Control Protocol}
\newacronym{tdd}{TDD}{Time Division Duplexing}
\newacronym{tdma}{TDMA}{Time Division Multiple Access}
\newacronym{tfl}{TfL}{Transport for London}
\newacronym{tm}{TM}{Transparent Mode}
\newacronym{trp}{TRP}{Transmitter Receiver Pair}
\newacronym{tti}{TTI}{Transmission Time Interval}
\newacronym{ttt}{TTT}{Time-to-Trigger}
\newacronym{tx}{TX}{Transmitter}
\newacronym{ue}{UE}{User Equipment}
\newacronym{ul}{UL}{Uplink}
\newacronym{uml}{UML}{Unified Modeling Language}
\newacronym{um}{UM}{Unacknowledged Mode}
\newacronym{utc}{UTC}{Urban Traffic Control}
\newacronym{vm}{VM}{Virtual Machine}
\newacronym{rsrq}{RSRQ}{Reference Signal Received Quality}
\newacronym{rssi}{RSSI}{Received Signal Strength Indicator}
\newacronym{crs}{CRS}{Cell Reference Signal}
\newacronym{nsa}{NSA}{Non Stand Alone}
\newacronym{mrdc}{MR-DC}{Multi \gls{rat} \gls{dc}}
\newacronym{endc}{EN-DC}{E-UTRAN-\gls{nr} \gls{dc}}
\newacronym{5gc}{5GC}{5G Core}
\newacronym{si}{SI}{Study Item}
\newacronym{iab}{IAB}{Integrated Access and Backhaul}
\newacronym{wf}{WF}{Wired-first}
\newacronym{hqf}{HQF}{Highest-quality-first}
\newacronym{pa}{PA}{Position-aware}
\newacronym{mlr}{MLR}{Maximum-local-rate}
\newacronym{wbf}{WBF}{Wired Bias Function}
\newacronym{mib}{MIB}{Master Information Block}
\newacronym{sib}{SIB}{Secondary Information Block}
\newacronym{kpi}{KPI}{Key Performance Indicator}
\newacronym{ppp}{PPP}{Poisson Point Process}
\newacronym{vr}{VR}{Virtual Reality}
\newacronym{scm}{SCM}{Spatial Channel Model}
\tikzstyle{startstop} = [rectangle, rounded corners, minimum width=2cm, minimum height=0.5cm,text centered, draw=black]
\tikzstyle{io} = [trapezium, trapezium left angle=70, trapezium right angle=110, minimum width=3cm, minimum height=1cm, text centered, draw=black]
\tikzstyle{process} = [rectangle, minimum width=2cm, minimum height=0.5cm, text centered, draw=black, alignb=center]
\tikzstyle{decision} = [ellipse, minimum width=2cm, minimum height=1cm, text centered, draw=black]
\tikzstyle{arrow} = [thick,<->,>=stealth]
\tikzstyle{line} = [thick,>=stealth]
\tikzstyle{darrow} = [thick,<->,>=stealth,dashed]
\tikzstyle{sarrow} = [thick,->,>=stealth]
\tikzstyle{larrow} = [line width=0.1mm,dashdotted,->,>=stealth]

\makeatletter
\def\grd@save@target#1{%
  \def\grd@target{#1}}
\def\grd@save@start#1{%
  \def\grd@start{#1}}
\tikzset{
  grid with coordinates/.style={
    to path={%
      \pgfextra{%
        \edef\grd@@target{(\tikztotarget)}%
        \tikz@scan@one@point\grd@save@target\grd@@target\relax
        \edef\grd@@start{(\tikztostart)}%
        \tikz@scan@one@point\grd@save@start\grd@@start\relax
        \draw[minor help lines] (\tikztostart) grid (\tikztotarget);
        \draw[major help lines] (\tikztostart) grid (\tikztotarget);
        \grd@start
        \pgfmathsetmacro{\grd@xa}{\the\pgf@x/1cm}
        \pgfmathsetmacro{\grd@ya}{\the\pgf@y/1cm}
        \grd@target
        \pgfmathsetmacro{\grd@xb}{\the\pgf@x/1cm}
        \pgfmathsetmacro{\grd@yb}{\the\pgf@y/1cm}
        \pgfmathsetmacro{\grd@xc}{\grd@xa + \pgfkeysvalueof{/tikz/grid with coordinates/major step x}}
        \pgfmathsetmacro{\grd@yc}{\grd@ya + \pgfkeysvalueof{/tikz/grid with coordinates/major step y}}
        \foreach \x in {\grd@xa,\grd@xc,...,\grd@xb}
        \node[anchor=north] at (\x,\grd@ya) {\pgfmathprintnumber{\x}};
        \foreach \y in {\grd@ya,\grd@yc,...,\grd@yb}
        \node[anchor=east] at (\grd@xa,\y) {\pgfmathprintnumber{\y}};
      }
    }
  },
  minor help lines/.style={
    help lines,
    gray,
    line cap =round,
    xstep=\pgfkeysvalueof{/tikz/grid with coordinates/minor step x},
    ystep=\pgfkeysvalueof{/tikz/grid with coordinates/minor step y}
  },
  major help lines/.style={
    help lines,
    line cap =round,
    line width=\pgfkeysvalueof{/tikz/grid with coordinates/major line width},
    xstep=\pgfkeysvalueof{/tikz/grid with coordinates/major step x},
    ystep=\pgfkeysvalueof{/tikz/grid with coordinates/major step y}
  },
  grid with coordinates/.cd,
  minor step x/.initial=.5,
  minor step y/.initial=.2,
  major step x/.initial=1,
  major step y/.initial=1,
  major line width/.initial=1pt,
}
\makeatother

\def\3G{\mathsf{3GPP}}

\definecolor{desireRed}{RGB}{230,57,60}%
\definecolor{darkPurple}{RGB}{59,31,43}%
\definecolor{springGreen}{RGB}{37,223,145}%
\definecolor{queenBlue}{RGB}{69,123,157}%
\definecolor{spaceCadet}{RGB}{29,53,87}%

\makeglossaries
\IEEEoverridecommandlockouts

\begin{document}
\glsunset{nr}

\title{Multi-Sector and Multi-Panel Performance\\in 5G mmWave Cellular Networks\vspace{-.3cm}}

\author{\IEEEauthorblockN{Mattia Rebato, Michele Polese, Michele Zorzi}
\IEEEauthorblockA{
\small University of Padova, Italy}
\small email:\texttt{\{rebatoma,polesemi,zorzi\}@dei.unipd.it}
\thanks{This work was partially supported by the U.S. Department of Commerce through NIST (Award No. 70NANB17H166).}}

\makeatletter
\patchcmd{\@maketitle}
  {\addvspace{0.5\baselineskip}\egroup}
  {\addvspace{-1\baselineskip}\egroup}
  {}
  {}
\makeatother

\flushbottom
\setlength{\parskip}{0ex plus0.1ex}

\maketitle

\begin{abstract}
The next generation of cellular networks (5G) will exploit the mmWave spectrum to increase the available capacity. Communication at such high frequencies, however, suffers from high path loss and blockage, therefore directional transmissions using antenna arrays and dense deployments are needed. Thus, when evaluating the performance of mmWave mobile networks, it is necessary to accurately model the complex channel, the directionality of the transmission, but also the interplay that these elements can have with the whole protocol stack, both in the radio access and in the higher layers. In this paper, we improve the channel model abstraction of the mmWave module for ns-3, by introducing the support of a more realistic antenna array model, compliant with 3GPP NR requirements, and of multiple antenna arrays at the base stations and mobile handsets. We then study the end-to-end performance of a mmWave cellular network by varying the channel and antenna array configurations, and show that increasing the number of antenna arrays and, consequently, the number of sectors is beneficial for both throughput and latency. 
\end{abstract}

\glsresetall
\glsunset{nr}

\begin{IEEEkeywords}
5G, millimeter wave, performance evaluation, beamforming, 3GPP, NR.
\end{IEEEkeywords}
\begin{picture}(0,0)(0,-345)
\centering
\put(0,0){
\put(-28,10){ M. Rebato, M. Polese, and M. Zorzi, ``Multi-Sector and Multi-Panel Performance in 5G mmWave Cellular Networks", in IEEE Global}
\put(-19,-0){Communications Conference: Communication QoS, Reliability and Modeling (Globecom2018 CQRM), Abu Dhabi, UAE, Dec 2018.}}
\end{picture}

\section{Introduction}
\label{sec:intro}
The next generation of cellular networks will need to cope with an ultra-high mobile traffic demand, due to the expected increase in the number of connected devices and to multimedia applications such as video streaming and \gls{vr}~\cite{cisco2017}. A possible enabler for these capacity-intensive applications is the communication in the mmWave band, i.e., approximately between 10 and 300~GHz, thanks to the availability of wide portions of free spectrum~\cite{rangan2017potentials}. Therefore, the fifth generation of cellular networks (5G), which is currently being standardized by the \gls{3gpp} as \gls{nr}\footnote{According to the latest 3GPP specifications~\cite{38300}, the acronym NR is used to refer to the 5G \acrlong{ran}.}, will exploit carrier frequencies in the mmWave spectrum, up to 52.6~GHz~\cite{38300}.

The \gls{phy} and \gls{mac} layer specifications for \gls{3gpp} \gls{nr} also include distinct procedures aimed at overcoming the main limitations of mmWave communications in a mobile environment~\cite{38300,38211}. The propagation at such high frequencies, indeed, suffers from a high path loss, which is proportional to the square of the carrier frequency.
A possible solution is the usage of directional communications, which are supported by \gls{3gpp} NR.
Antenna arrays with a large number of elements can be used to generate narrow beams and increase the link budget with the beamforming gain~\cite{rangan2017potentials}.
Given the small wavelength at mmWaves, it is possible to pack many antenna elements in a small area: for example, at 30~GHz the wavelength $\lambda$ is approximately 1~cm, thus a rectangular array with 16 antennas (4 by 4) spaced by $\lambda/2$ would fit in a package with area smaller than a 2~cm by 2~cm square, and can be installed into a modern smartphone or \gls{vr} headset. 

Additionally, mmWave signals are easily blocked by common materials such as brick, mortar, or even by the human body~\cite{lu2012modeling}, thus the quality of the mmWave signal can exhibit high variability over time, with variations in the received power in the order of 30~dB for transitions between \gls{los} and \gls{nlos}. Therefore, \gls{nr} will use multi-connectivity solutions with a sub-6~GHz radio overlay for coverage, possibly based on \gls{lte} thanks to a tight internetworking~\cite{37340}, and mmWave links for capacity. 
Finally, the harsh propagation environment at mmWaves has an impact on the whole protocol stack, with higher layer protocols, such as those at the transport layer, being affected by the complex interplay with the mmWave channel variability~\cite{pieska2017tcp,polese2017tcp}.

\begin{figure}[t!]
\setlength\belowcaptionskip{0cm}
\centering
\includegraphics[width=\columnwidth]{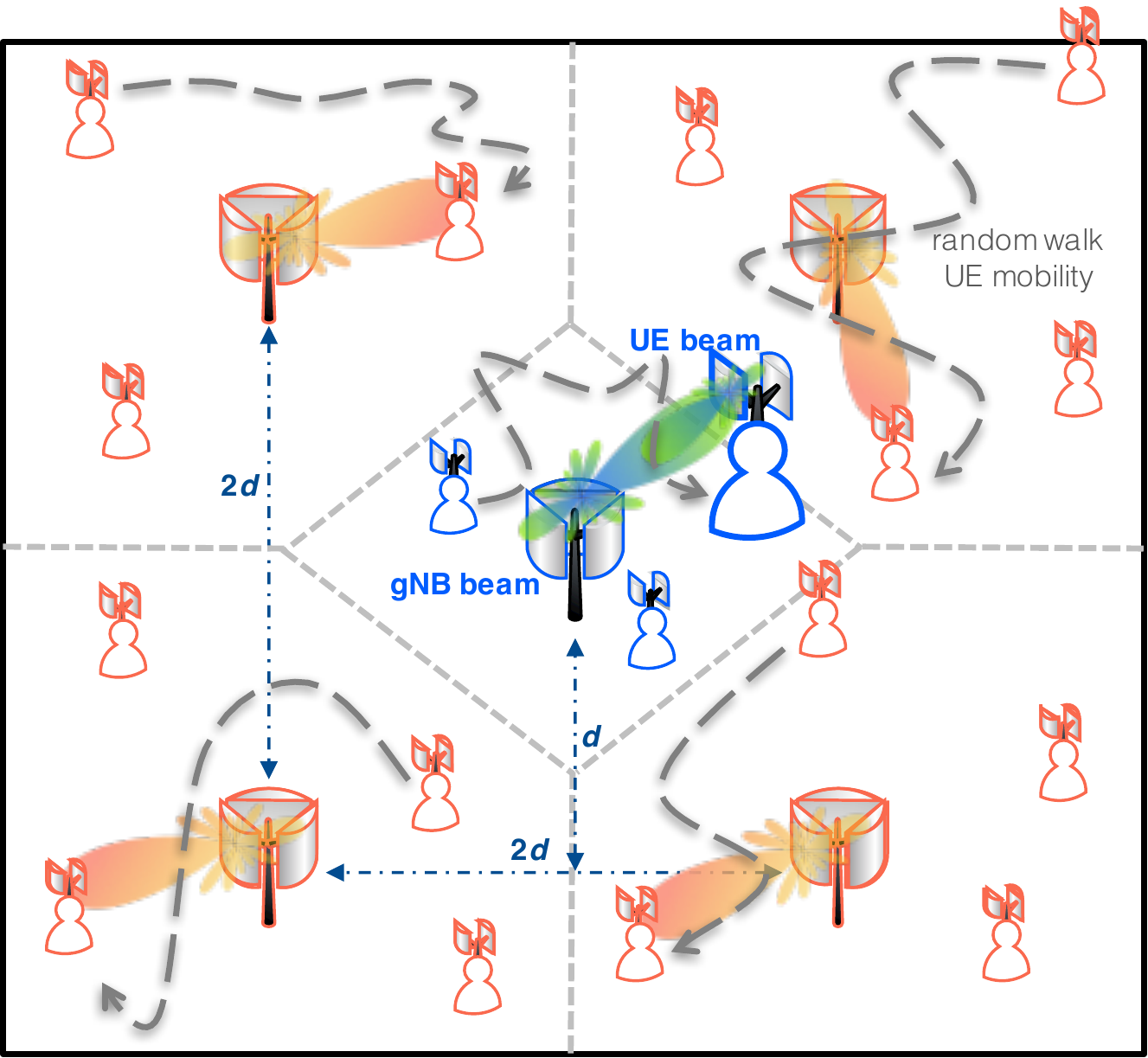}
\caption{Illustration of  a possible configuration for the scenario considered, where five \glspl{gnb} are placed in the area and are serving all the users using three sectors. Meanwhile, the \glspl{ue} are equipped with only two arrays and move randomly accordingly to a two-dimensional \emph{random walk model}.}
\vspace{-0.5cm}
\label{fig:upd_global_view}
\end{figure}
Given the complexity of the interactions between the underlying physical propagation phenomena and the full protocol stack, it is important to carefully consider every single component of an end-to-end mmWave cellular network when analyzing its performance. System level simulators are natural candidates for end-to-end performance evaluations, given that, as of today, there are no real deployments at scale of \gls{nr} mmWave cellular networks. In~\cite{mezzavilla2017end}, NYU and the University of Padova have introduced a mmWave cellular network module for the popular ns-3 simulator~\cite{henderson2008network}, which features the implementation of the \gls{3gpp} channel model for frequencies above 6~GHz~\cite{38900}, and a \gls{3gpp}-like cellular protocol stack.

In this paper we extend the model in~\cite{mezzavilla2017end}, introducing the possibility of deploying multiple antenna arrays at each \gls{gnb} and \gls{ue}, thus allowing a sectorized deployment.
Moreover, we add the possibility of simulating non-isotropic antenna patterns for each single antenna elements in each array, following the \gls{3gpp} specifications. 
With the aim to provide an accurate tool to evaluate the end-to-end performance in mmWave mobile scenarios, our framework simulates the radiation pattern of a patch antenna element and provides the precise antenna gain in each direction.
Moreover, we also report simulation results based on the realistic antenna pattern extension, considering end-to-end transport protocols (i.e., UDP) and a multi-site cellular network deployment. In particular, we characterize the system throughput and latency when varying the number of antenna arrays at the \gls{gnb}, as well as the user and \gls{gnb} density, and the \gls{3gpp} deployment scenario. We show that by increasing the number of sectors it is possible to improve the \gls{sinr}, thus increasing throughput and decreasing latency, especially for the users with the worst channel conditions.
%\michelep{more on results?}.

The rest of the paper is organized as follows. In Sec.~\ref{sec:bf} we provide insights on the \gls{3gpp} antenna array model, with references to the standards, and detail its ns-3 implementation. In Sec.~\ref{sec:perf} we describe our simulation setup, and report the results of the simulation campaign. Finally, in Sec.~\ref{sec:concl} we conclude the paper and suggest possible extensions.

\section{3GPP Realistic Antenna Array Model}
\label{sec:bf}
The ns-3 mmWave module provides an implementation of the \gls{3gpp} channel model for frequencies above 6~GHz~\cite{38900,zhang2017ns3}, which should be used for the evaluation of \gls{nr} networks at mmWave frequencies. 
It is a \gls{scm}, i.e., the channel is represented by a matrix $\mathbf{H}$, whose entry $(t,r)$ models the channel between the $t$-th and the $r$-th antenna elements at the transmitter and the receiver, respectively. Each entry $(t,r)$ is given by the contribution of $N$ clusters, which represent the direct \gls{los} path (if present) and the additional \gls{nlos} reflections. Each cluster is modeled using different powers and delays, and depends itself on multiple rays, distributed around a common cluster angle of arrival and departure. 

The implementation of the ns-3 mmWave module also includes an optimal beamforming model, which assumes a perfect knowledge of the channel matrix $\mathbf{H}$, and a simple brute-force beam search method. Nevertheless, the implementation described in~\cite{zhang2017ns3} supports a single panel per \gls{gnb} (or \gls{ue}), with isotropic antenna elements. However, two assumptions of the available models have a limited realism. First, having a perfect knowledge of the channel is not feasible in practice, even though a partial estimation can be achieved using reference and synchronization signals. Second, also the use of antenna elements with an isotropic radiation pattern is a non-realistic hypothesis, given that such antenna elements do not exist~\cite{balanis}.

In this paper, we introduce in the ns-3 mmWave module the support for the \gls{3gpp} antenna array model, which makes it possible to precisely determine the array radiation pattern as suggested in \gls{3gpp} Technical Reports~\cite{38900,37840,36873}, and does not use isotropic antenna elements. Moreover, given that with non-isotropic elements it is not possible to uniformly cover the whole angular space, we extend the channel model classes with a multi-sector model for the \glspl{gnb}, with each sector covered by a different antenna array, and with a multi-panel model for \glspl{ue}\footnote{
	Using the \gls{3gpp} terminology, at the \gls{ue} multiple arrays are associated to different \emph{panels}. At the \gls{gnb}, instead, each array covers a \emph{sector}.
	}. 

In this section, we first provide a mathematical characterization of the antenna array patterns and of the field factor $\mathbf{F}$, and then describe how the realistic antenna model was implemented in ns-3.

% We briefly describe in this section how to compute the realistic antenna array pattern considered, and report expressions for its field factors $F$. The field factor makes it possible to integrate the array pattern into the channel matrix $\mathbf{H}$.
% In this manner, we can model the effect of the radiation pattern in each ray handled in the channel.

\subsection{Antenna pattern definition}
\label{antenna_pattern_computation}

The radiation pattern of the whole array, defined also as the \emph{array radiation pattern} $A_A$, is given by the superposition of its array factor $\mathsf{AF}$, which models the directivity of an antenna array, and the element radiation pattern $A_E$.
The latter takes into account how power is radiated by the single antenna elements~\cite{rebato18}.

Let us first consider the element radiation pattern $A_E$, which characterizes how the power is radiated by a single antenna element in all possible directions. It is defined for any pair of vertical and horizontal angles $(\theta,\phi)$, and is fundamental in scenarios where directional transmission is used, because it precisely models the direction through which the antenna element transmits or receives power.
Following the \gls{3gpp} specifications, the $A_E$ of each single antenna element is composed of horizontal and vertical radiation patterns~\cite{38900,36873}.
Specifically, the latter, $A_{E,V}(\theta)$, is obtained as
\begin{equation}
A_{E,V}(\theta) = - \min \left\{ 12 \left( \frac{\theta - 90}{\theta_{3 \text{dB}}}\right)^2, SLA_V \right\},
\end{equation}
where $\theta_{3 \text{dB}}$ is the vertical 3~dB beamwidth, and $SLA_V = 30 \text{ dB}$ is the side-lobe level limit.
Similarly, the horizontal pattern is computed as 
\begin{equation}
A_{E,H}(\phi) = - \min \left\{ 12 \left( \frac{\phi}{\phi_{3 \text{dB}}}\right)^2, A_m \right\},
\end{equation}
where $\phi_{3 \text{dB}}$ is the horizontal 3~dB beamwidth, and $ A_m = 30$~dB is the front-back ratio.
By considering both the vertical and horizontal patterns it is possible to obtain the 3D antenna element gain for each angular direction as
\begin{equation}
A_E(\theta,\phi) = G_{\max}- \min \left\{- \left[ A_{E,V}(\theta) + A_{E,H}(\phi) \right], A_m \right\},
\label{global_pattern_equation}
\end{equation}
where $G_{\max}$ is the maximum directional gain in the main-lobe direction of the antenna element~\cite{38900,rebato18}. 
The expression in Eq.~\eqref{global_pattern_equation} provides the gain in dB that can be applied to a single ray of a cluster, with angle $(\theta,\phi)$, due to the effect of the element radiation pattern.
Notice that some of these antenna settings, such as the directivity $G_{\max}$ and the 3 dB beamwidths $\phi_{3 \text{dB}}$ and $\theta_{3 \text{dB}}$, differ in the \glspl{gnb} and the \glspl{ue}.
For this reason, we report in Tab.~\ref{table_3gpp_antenna_settings} the values of the different parameters for the \gls{gnb} and UE antennas, as suggested in the \gls{3gpp} specifications.
The values in Tab.~\ref{table_3gpp_antenna_settings} will be used in our performance evaluation.  

%Using the definition of the element radiation pattern in Eq.~\eqref{global_pattern_equation}, we can define expression of the array radiation pattern.  
Consequently, the radiation of the entire array is obtained considering the effect of all the single elements, through the radiation pattern $A_E(\theta,\phi)$, and is defined, following~\cite{37840}, as
\begin{equation}
A_A(\theta,\phi) = A_E(\theta,\phi) + \mathsf{AF}(\theta,\phi).
\label{ralation_array_element}
\end{equation}
This last equation considers the effect of the element radiation pattern in combination with the array factor $\mathsf{AF}(\theta,\phi)$.
Further details regarding all the previously reported antenna terms and expressions can be found in~\cite{37840,rebato18}.

Finally, in order to apply the array radiation pattern $A_A$ to the channel matrix $\textbf{H}$, we compute the field pattern $\mathbf{F}$~\cite{36873}, which is composed by a vertical and a horizontal polarization term, i.e., 
\begin{align}
\begin{cases}
F_{\theta} (\theta,\phi) &= \sqrt{A_A(\theta,\phi)}\cos (\zeta),\\
F_{\phi } (\theta,\phi) &= \sqrt{A_A(\theta,\phi)}\sin (\zeta),
\end{cases}\\
\mathbf{F}(\theta,\phi) = \left[ F_{\theta} (\theta,\phi), F_{\phi} (\theta,\phi)\right]
\label{eq:field_pattern}
\end{align} 
respectively, where $\zeta$ is the polarization slant angle and $A_A(\theta,\phi)$ is the 3D antenna array gain pattern previously obtained in Eq.~\eqref{ralation_array_element}.

\begin{table}[t]
\centering
\vspace{0.1in}
\caption{\gls{gnb} and UE suggested settings from~\cite{38802}. Moreover, vertical and horizontal spacing of antenna elements ($dy,dz$) is kept equal and fixed to $0.5\lambda$ for both \gls{gnb} and UE.}
\label{table_3gpp_antenna_settings}
\begin{tabular}{r|c|c|c}
\toprule
    & \begin{tabular}[c]{@{}c@{}}directivity\\ $G_{\max}$\end{tabular} & \begin{tabular}[c]{@{}c@{}}HPBW\\
    $(\theta_{3 \text{dB}},\phi_{3 \text{dB}})$\end{tabular} & \# sectors/panels \\ \hline
	\gls{gnb} & 8 dBi  & ($65^\circ,65^\circ$) & 3  \\
	\gls{ue}  & 5 dBi & ($90^\circ,90^\circ$) & 2  \\
\bottomrule
\end{tabular}
\end{table}
According to the \gls{3gpp} channel definition (Equation (7.5-22) in~\cite{38900}) the field factor can be easily considered in the channel matrix and, with a slight abuse of notation, each element of the channel matrix $\textbf{H}$, for a single cluster, can be represented as
\begin{equation}
h_{r,t}= \sum_{m=1}^{M} \left[\mathbf{F}_{r}\left(\Omega^{r}_{m}\right)\right]^T \mathbf{g}_{m} \mathbf{F}_{t}\left(\Omega^{t}_{m}\right) u_{r}\left(\Omega^{r}_{m}\right) u^*_{t}\left(\Omega^{t}_{m}\right),
\label{eq:channel_matrix}
\end{equation}
where $t$ and $r$ are the indices of the $t$-th and $r$-th elements of the transmitter and receiver array, respectively, $\mathbf{g}_{m}$ is the small-scale fading gain of ray $m$, $\mathbf{F}_{r}$ and $\mathbf{F}_{t}$ are the receiver and transmitter field patterns previously computed in Eq.~\eqref{eq:field_pattern}, and $u_{r}(\cdot)$ and $u_{t}(\cdot)$ indicate the 3D spatial signature elements of the receiver and transmitter, respectively.
Moreover, $\Omega^{r}_{m} = (\theta_m^r,\phi_m^r)$ is the angular spread of the vertical and horizontal angles of arrival and $\Omega^{t}_{m} = (\theta_m^t,\phi_m^t)$ is the angular spread of the vertical and horizontal angles of departure.

\begin{figure}[t]
	\centering
	\setlength\belowcaptionskip{-.6cm}
	\includegraphics[width=.9\columnwidth]{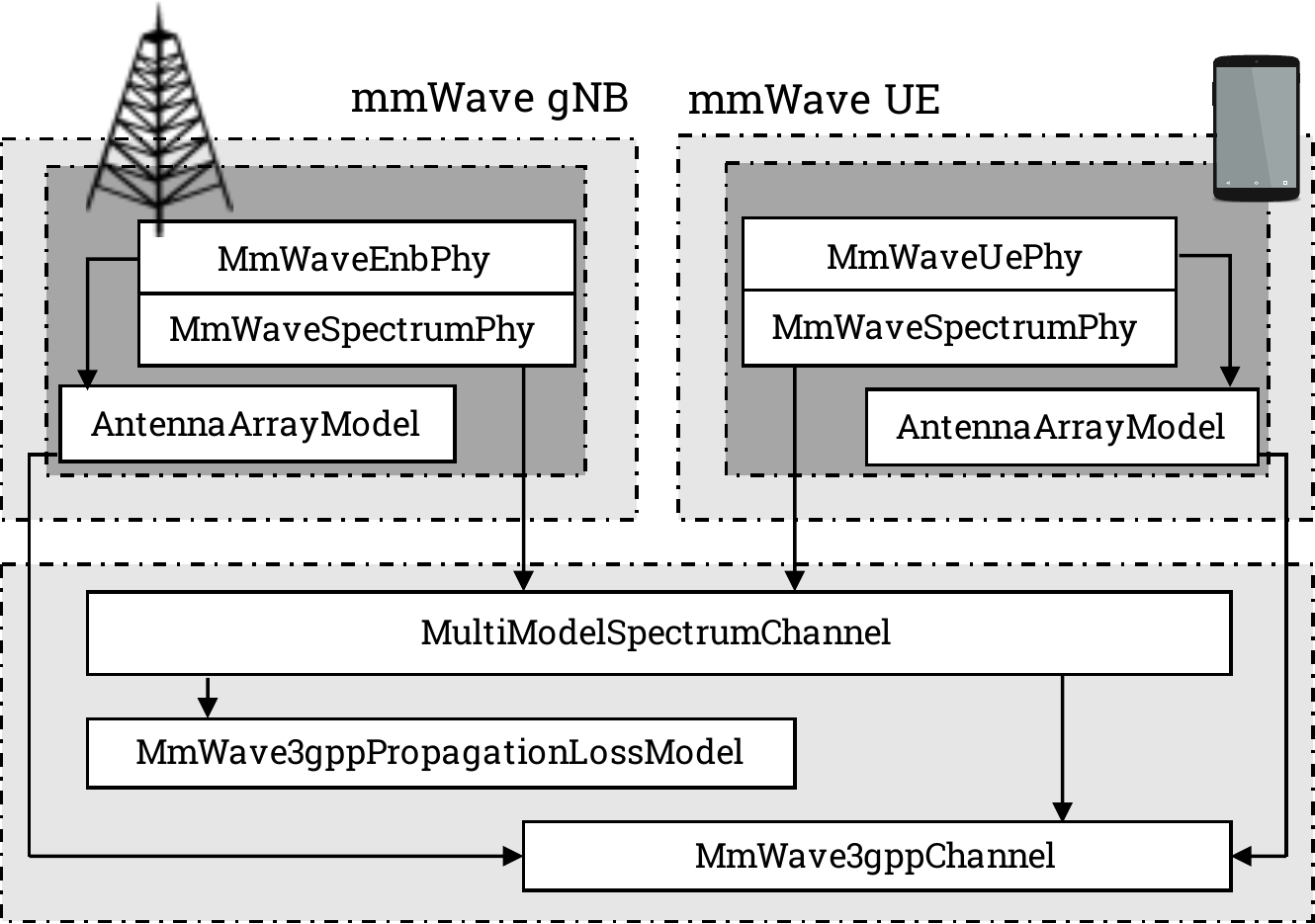}
	\caption{\gls{phy} layer and channel modeling in the ns-3 mmWave framework.}
	\label{fig:ns3stack}
\end{figure}

\subsection{ns-3 integration}
\label{ns3_integration}

As already mentioned, in this work we extend the ns-3 framework in~\cite{mezzavilla2017end} by incorporating the possibility to simulate the realistic antenna patterns and configurations described in the previous section. As shown in Fig.~\ref{fig:ns3stack}, the channel model implementation in the ns-3 mmWave module depends on a number of classes, with different functionalities. The propagation loss is computed by the \texttt{MmWave3gppPropagationLossModel} class, which implements also a probabilistic model for the \gls{los} and \gls{nlos} condition according to~\cite{38900}. The \texttt{MmWave3gppChannel}, instead, computes the channel matrix $\mathbf{H}$ for each single transmitter-receiver pair, and applies the beamforming vectors to get the beamformed received power spectral density. Moreover, the \texttt{AntennaArrayModel} class models the antenna arrays at the \gls{gnb} and the \gls{ue}. Finally, in each terminal an instance of the \texttt{MmWaveSpectrumPhy} handles the interaction between the \gls{phy} layer implementation, the error model and the channel abstraction.

In order to implement the \gls{3gpp} antenna array model, the \texttt{AntennaArrayModel} class has been extended to properly handle the presence of multiple antenna arrays, allowing each terminal to transmit and receive with the proper sector (or panel) according to the angular direction of the other transceiver in the link. 
In the same class, we introduce the possibility of modeling accurate antenna radiation patterns, providing the field patterns $\mathbf{F}$ to the \texttt{MmWave3gppChannel}, which applies them to the channel matrix. In this first version, we consider the \gls{los} direction to compute the beamforming vector pair for the link.
Future extensions will include the possibility of performing a codebook-based beamforming, with a realistic cell scan.

We highlight that all the introduced antenna settings are tunable using the ns-3 \emph{attributes} system, therefore the framework can be adjusted to simulate \gls{3gpp} \gls{nr} specifications (i.e., with the settings in Tab.~\ref{table_3gpp_antenna_settings}), but also other configurations, resulting in a useful tool for the evaluation of realistic end-to-end mmWave networks.

\vspace{-0.3cm}
\section{Performance Evaluation}
\label{sec:perf}
\subsection{Scenario}
In this paper, we study the performance in terms of end-to-end user throughput and latency in a multi-site deployment, using UDP as the transport protocol. We consider the scenario in Fig.~\ref{fig:upd_global_view}, with 4~\glspl{gnb} at the vertices of a square and a \gls{gnb} at the center. We consider the distance $d$ as a parameter, which can vary in $d\in[100,200]$~m, a typical range for an ultra-dense mmWave small cell deployment. In this setup, the 5G network is deployed in a \gls{nsa} mode, i.e., it uses a 4G \gls{epc} network, and the \glspl{ue} are configured with multi-connectivity between an \gls{lte} \gls{enb} (co-deployed with the central \gls{gnb}) and an \gls{nr} \gls{gnb}~\cite{37340}. The end-to-end flows are configured as split bearers, i.e., the \gls{lte} \gls{enb} acts as a local traffic anchor with respect to the core network, and data packets are forwarded to and from the mmWave \glspl{gnb}, according to the configuration described in~\cite{polese2016performance,polese2017jsac}. All the base stations are interconnected with X2 links, which are realistically simulated in terms of data rate limit and additional latency, as shown by the parameters in Tab.~\ref{table:params}.

There are $N_{UE} =$ 25 or 50 users in the scenario, and they move randomly according to a two-dimensional random walk model. They can freely hand over between the different mmWave \glspl{gnb}, or switch to the \gls{lte} \gls{enb} if all the mmWave links are in the outage condition (i.e., with an \gls{sinr} below $-5$ dB). The handover procedure is coordinated by the central \gls{lte} unit, and avoids latency-consuming interactions with the core network, as described in~\cite{polese2017jsac}. The users consume content from a remote server (e.g., for video streaming), with a constant bitrate $R_{UE} = 100$ Mbit/s. 
\begin{table}[t]
  \setlength{\belowcaptionskip}{-0.2cm}
  \vspace{0.05in}
  \centering
    \caption{Additional simulation parameters.}
  \label{table:params}
  \small
  \begin{tabular}{@{}ll@{}}
    \toprule
    Parameter & Value \\
    \midrule
    mmWave carrier frequency & 28 GHz \\
    mmWave bandwidth & 1 GHz \\
    \gls{3gpp} Channel Scenario & Urban Micro, Urban Macro\\
    mmWave outage threshold $\Omega$ & $- 5$ dB \\
    mmWave max PHY rate & 3.2 Gbit/s \\
    X2 link latency $D_{X2}$  & 1 ms \\
    S1 link latency $D_{S1}$  & 10 ms \\
    % \acrshort{pgw} to remote server latency $D_{RS}$ & $[0, 10, 20]$~ms\\
    RLC buffer size $B_{RLC}$ & 5 MB \\
    RLC AM reordering timer & 1 ms \\
    S1-MME link latency $D_{MME}$ & 10 ms \\
    UE speed $v$ & $\mathcal{U}[2,4]$ m/s \\
    UDP source rate $R_{UE}$ & 100 Mbit/s \\
    \bottomrule
  \end{tabular}
  \vspace{-.3cm}
\end{table}
We test a different number of sectors for each mmWave \gls{gnb}, ranging from 3 to 4. In this first evaluation, we do not consider the single-sector setup with isotropic antenna elements that is still available in the ns-3 mmWave module, since it would be less realistic. The antenna directivity in our simulations is configured as described in Sec.~\ref{sec:bf}. The \glspl{ue} are equipped with 2 panels~\cite{38802}.
We also compare the results for two different \gls{3gpp} channel model configurations, namely the Urban Macro (UMa) and the Urban Micro (UMi) scenarios. The channel condition between each user and each \gls{gnb} is randomly assigned according to the \gls{3gpp} model~\cite{38900}. 

The metrics we consider are the end-to-end throughput, measured above the transport layer for each user, and the latency in the \gls{ran}. In particular, in our simulation setup, the end-to-end latency is given by a fixed component in the wired part of the connection, and by a variable one in the \gls{ran} (i.e., the PDCP layer latency), which depends on the different configurations we examine. Therefore, in the following sections we will only report the PDCP layer latency. 

\begin{figure}[t]
	\centering
	\setlength\belowcaptionskip{-.3cm}
	\setlength\fwidth{0.95\columnwidth}    
	\setlength\fheight{0.55\columnwidth}
% \definecolor{mycolor1}{rgb}{0.20810,0.16630,0.52920}%
% \definecolor{mycolor2}{rgb}{0.21783,0.72504,0.61926}%
% \definecolor{mycolor3}{rgb}{0.97630,0.98310,0.05380}%
\definecolor{mycolor1}{rgb}{0.00000,0.44700,0.74100}%
\definecolor{mycolor4}{rgb}{0.6784,0.8471,0.9020}%
\definecolor{mycolor2}{rgb}{0.85000,0.32500,0.09800}%
\definecolor{mycolor5}{rgb}{1.0000,0.6275,0.4784}%
\definecolor{mycolor3}{rgb}{0.92900,0.69400,0.12500}%
\definecolor{mycolor6}{rgb}{1.0000,0.9804,0.8039}%
\begin{tikzpicture}
\pgfplotsset{every tick label/.append style={font=\scriptsize}}

\begin{axis}[%
width=0.951\fwidth,
height=\fheight,
at={(0\fwidth,0\fheight)},
scale only axis,
bar shift auto,
xmin=0,
xmax=6,
xtick={0,0.8,2,4,5.2,6},
xticklabel style={text width=50, align=center, font=\scriptsize},
xticklabels={{},{$d=100$~m $N_{UE} = 25$},{$d=200$~m $N_{UE} = 25$},{$d=100$~m $N_{UE} = 50$},{$d=200$~m $N_{UE} = 50$},{}},
ymin=0,
ymax=100,
xlabel style={font=\scriptsize\color{white!15!black}},
xlabel={Scenario configuration},
ylabel style={font=\scriptsize\color{white!15!black}},
ylabel={Average UDP user throughput [Mbit/s]},
axis background/.style={fill=white},
ylabel shift = -5 pt,
yticklabel shift = -2 pt,
% xlabel shift = -4 pt,
xticklabel shift = -1 pt,
axis x line*=bottom,
axis y line*=left,
xmajorgrids,
ymajorgrids,
legend style={font=\scriptsize,legend cell align=left,align=left,draw=white!15!black, at={(0.99,0.99)},anchor=north east},
]
%  
% \addplot[ybar, xshift=-0.1cm, opacity=1, postaction={pattern=dots}, bar width=0.2, fill=mycolor1, draw=black, area legend] table[row sep=crcr] {%
% 0.8	94.2812678587298\\
% 2	74.6122691474347\\
% 3	0\\
% 4	78.5535442468359\\
% 5.2	62.1687832447913\\
% };
% \addplot[forget plot, color=white!15!black] table[row sep=crcr] {%
% 0	0\\
% 6	0\\
% };
%\addlegendentry{2 panels}

%  
\addplot[ybar, xshift=-0.1cm, opacity=1, postaction={pattern=dots}, bar width=0.2, fill=mycolor2, draw=black, area legend] table[row sep=crcr] {%
0.8	94.8310223164927\\
2	77.6375984724848\\
3	0\\
4	78.5763544422887\\
5.2	61.6399866508433\\
};
\addplot[forget plot, color=white!15!black] table[row sep=crcr] {%
0	0\\
6	0\\
};
%\addlegendentry{3 panels}

%  
\addplot[ybar, xshift=-0.1cm, opacity=1, postaction={pattern=dots}, bar width=0.2, fill=mycolor3, draw=black, area legend] table[row sep=crcr] {%
0.8	95.4968099548586\\
2	79.4634071680163\\
3	0\\
4	79.8799340317218\\
5.2	61.7740751901941\\
};
\addplot[forget plot, color=white!15!black] table[row sep=crcr] {%
0	0\\
6	0\\
};
%\addlegendentry{4 panels}

\end{axis}

\begin{axis}[%
width=0.951\fwidth,
height=\fheight,
at={(0\fwidth,0\fheight)},
axis x line=none,%axis on top,
hide x axis,
axis line style={-},
scale only axis,
bar shift auto,
xmin=0,
xmax=6,
xtick={0,0.8,2,4,5.2,6},
ymin=0,
ymax=100,
hide y axis,
% axis background/.style={fill=white},
legend style={font=\scriptsize,legend cell align=left,align=left,draw=white!15!black, at={(0.99,0.99)},anchor=north east},
]
% \addplot[ybar, bar width=0.2, opacity=1, fill=mycolor1, draw=black, area legend] table[row sep=crcr] {%
% 0.8	90.749504640899\\
% 2	73.9652082150478\\
% 3	0\\
% 4	74.2850479009646\\
% 5.2	59.4727809869225\\
% };
% \addplot[forget plot, color=white!15!black] table[row sep=crcr] {%
% 0	0\\
% 6	0\\
% };
% \addlegendentry{2 sectors}

\addplot[ybar, bar width=0.2, opacity=1, fill=mycolor2, draw=black, area legend] table[row sep=crcr] {%
0.8	90.9073688089458\\
2	74.4472226576182\\
3	0\\
4	73.5720508595527\\
5.2	59.5390929823385\\
};
\addplot[forget plot, color=white!15!black] table[row sep=crcr] {%
0	0\\
6	0\\
};
\addlegendentry{3 sectors}

\addplot[ybar, bar width=0.2, opacity=1, fill=mycolor3, draw=black, area legend] table[row sep=crcr] {%
0.8	92.5002541975188\\
2	77.3386230249862\\
3	0\\
4	75.8228193917921\\
5.2	62.176485855643\\
};
\addplot[forget plot, color=white!15!black] table[row sep=crcr] {%
0	0\\
6	0\\
};
\addlegendentry{4 sectors}

\end{axis}

\pgfplotsset{ticks=none}
\begin{axis}[%
ybar,
width=0.951\fwidth,
height=\fheight,
at={(0\fwidth,0\fheight)},
scale only axis,
xtick=data,
%xmax=3000,
%xlabel style={font=\scriptsize\color{white!15!black}},
%xlabel={Distance [m]},
%xmajorgrids,
ymin=0,
ymax=100,
%ylabel style={font=\scriptsize\color{white!15!black}},
%ylabel={End-to-end latency [ms]},
%ymajorgrids,
ylabel shift = -5 pt,
yticklabel shift = -2 pt,
legend style={font=\scriptsize,at={(0.5,0.99)},anchor=north,legend cell align=left,align=left,draw=white!15!black},
enlarge x limits=0.15,
hide y axis,
hide x axis,
legend columns=2,
]
\addplot [fill=white,postaction={pattern=crosshatch dots}]
  table[row sep=crcr]{%
  1 0\\
  2 0\\
  3 0\\
  4 0\\
};
\addlegendentry{UMi}

\addplot [fill=white]
  table[row sep=crcr]{%
  1 0\\
  2 0\\
  3 0\\
  4 0\\
};
\addlegendentry{UMa}
\end{axis}

\end{tikzpicture}%
    \caption{Average user throughput for different configurations of the sectors, distances $d$ and number of users $N_{UE}$.}
    \label{fig:throughput}
\end{figure}
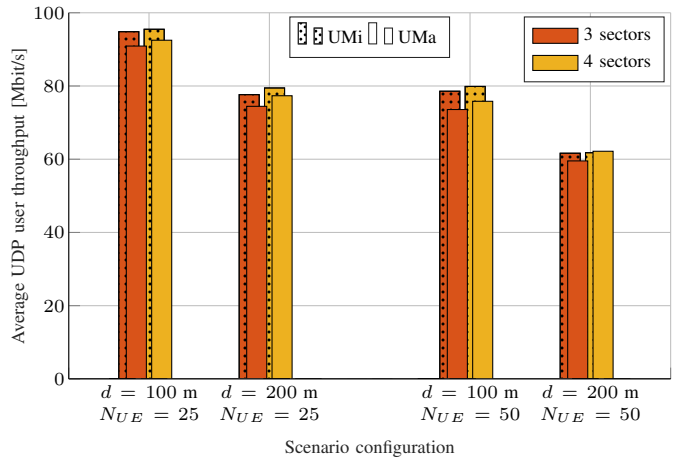

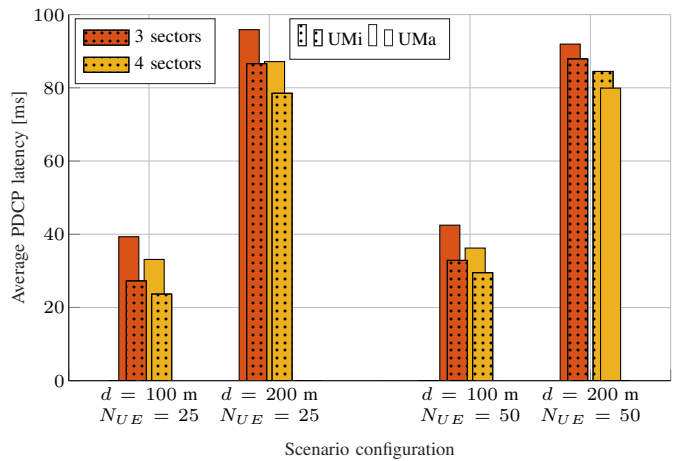
\begin{figure}[t]
	\centering
		\setlength\belowcaptionskip{-0.6cm}
	\setlength\fwidth{0.95\columnwidth}    
	\setlength\fheight{0.55\columnwidth}
% \definecolor{mycolor1}{rgb}{0.20810,0.16630,0.52920}%
% \definecolor{mycolor2}{rgb}{0.21783,0.72504,0.61926}%
% \definecolor{mycolor3}{rgb}{0.97630,0.98310,0.05380}%
\definecolor{mycolor1}{rgb}{0.00000,0.44700,0.74100}%
\definecolor{mycolor4}{rgb}{0.6784,0.8471,0.9020}%
\definecolor{mycolor2}{rgb}{0.85000,0.32500,0.09800}%
\definecolor{mycolor5}{rgb}{1.0000,0.6275,0.4784}%
\definecolor{mycolor3}{rgb}{0.92900,0.69400,0.12500}%
\definecolor{mycolor6}{rgb}{1.0000,0.9804,0.8039}%
\begin{tikzpicture}
\pgfplotsset{every tick label/.append style={font=\scriptsize}}

\begin{axis}[%
width=0.951\fwidth,
height=\fheight,
at={(0\fwidth,0\fheight)},
scale only axis,
bar shift auto,
xmin=0,
xmax=6,
xtick={0,0.8,2,4,5.2,6},
xticklabel style={text width=50, align=center, font=\scriptsize},
xticklabels={{},{$d=100$~m $N_{UE} = 25$},{$d=200$~m $N_{UE} = 25$},{$d=100$~m $N_{UE} = 50$},{$d=200$~m $N_{UE} = 50$},{}},
ymin=0,
ymax=100,
xlabel style={font=\scriptsize\color{white!15!black}},
xlabel={Scenario configuration},
ylabel style={font=\scriptsize\color{white!15!black}},
ylabel={Average PDCP latency [ms]},
axis background/.style={fill=white},
ylabel shift = -5 pt,
yticklabel shift = -2 pt,
% xlabel shift = -4 pt,
xticklabel shift = -1 pt,
axis x line*=bottom,
axis y line*=left,
xmajorgrids,
ymajorgrids,
legend style={font=\scriptsize,legend cell align=left,align=left,draw=white!15!black, at={(0.99,0.99)},anchor=north east},
]
% \addplot[ybar, xshift=-0.1cm, opacity=1, bar width=0.2, fill=mycolor1, draw=black, area legend] table[row sep=crcr] {%
% 0.8	41.2698740835829\\
% 2	97.3644619044938\\
% 3	0\\
% 4	42.1918265468762\\
% 5.2	89.3970814000509\\
% };
% \addplot[forget plot, color=white!15!black] table[row sep=crcr] {%
% 0	0\\
% 6	0\\
% };
% %\addlegendentry{2 panels}

\addplot[ybar, xshift=-0.1cm, opacity=1, bar width=0.2, fill=mycolor2, draw=black, area legend] table[row sep=crcr] {%
0.8	39.3267148030508\\
2	95.8764666002246\\
3	0\\
4	42.4870467349923\\
5.2	91.9382882212745\\
};
\addplot[forget plot, color=white!15!black] table[row sep=crcr] {%
0	0\\
6	0\\
};
%\addlegendentry{3 panels}

\addplot[ybar, xshift=-0.1cm, opacity=1, bar width=0.2, fill=mycolor3, draw=black, area legend] table[row sep=crcr] {%
0.8	33.1225084592977\\
2	87.1491617993407\\
3	0\\
4	36.2286814416412\\
5.2	0\\
};
\addplot[forget plot, color=white!15!black] table[row sep=crcr] {%
0	0\\
6	0\\
};
%\addlegendentry{4 panels}

\end{axis}

\begin{axis}[%
width=0.951\fwidth,
height=\fheight,
at={(0\fwidth,0\fheight)},
axis x line=none,%axis on top,
hide x axis,
axis line style={-},
scale only axis,
bar shift auto,
xmin=0,
xmax=6,
xtick={0,0.8,2,4,5.2,6},
ymin=0,
ymax=100,
hide y axis,
% axis background/.style={fill=white},
legend style={font=\scriptsize,legend cell align=left,align=left,draw=white!15!black, at={(0.01,0.99)},anchor=north west},
]
% \addplot[ybar, bar width=0.2, fill=mycolor1, postaction={pattern=dots}, draw=black, area legend] table[row sep=crcr] {%
% 0.8	27.3656634205145\\
% 2	97.1761146433689\\
% 3	0\\
% 4	33.3413409224255\\
% 5.2	89.5072221169529\\
% };
% \addplot[forget plot, color=white!15!black] table[row sep=crcr] {%
% 0	0\\
% 6	0\\
% };
% \addlegendentry{2 sectors}

\addplot[ybar, bar width=0.2, fill=mycolor2, postaction={pattern=dots}, draw=black, area legend] table[row sep=crcr] {%
0.8	27.2767434135122\\
2	86.6035725095101\\
3	0\\
4	32.8604611528982\\
5.2	87.8982674393984\\
};
\addplot[forget plot, color=white!15!black] table[row sep=crcr] {%
0	0\\
6	0\\
};
\addlegendentry{3 sectors}

\addplot[ybar, bar width=0.2, fill=mycolor3, postaction={pattern=dots}, draw=black, area legend] table[row sep=crcr] {%
0.8	23.6720623193619\\
2	78.5148747359346\\
3	0\\
4	29.5015994598762\\
5.2	84.4684883585893\\
};
\addplot[forget plot, color=white!15!black] table[row sep=crcr] {%
0	0\\
6	0\\
};
\addlegendentry{4 sectors}

\end{axis}

\begin{axis}[%
width=0.951\fwidth,
height=\fheight,
at={(0\fwidth,0\fheight)},
scale only axis,
bar shift auto,
xmin=0,
xmax=6,
xtick={0,0.8,2,4,5.2,6},
ymin=0,
ymax=100,
xlabel style={font=\scriptsize\color{white!15!black}},
ylabel style={font=\scriptsize\color{white!15!black}},
% axis background/.style={fill=white},
ylabel shift = -5 pt,
yticklabel shift = -2 pt,
% xlabel shift = -4 pt,
xticklabel shift = -1 pt,
axis x line=none,
axis y line=none,
]
% \addplot[ybar, xshift=-0.1cm, opacity=1, bar width=0.2, fill=mycolor1, draw=black, area legend] table[row sep=crcr] {%
% 0.8 41.2698740835829\\
% 2 97.3644619044938\\
% 3 0\\
% 4 42.1918265468762\\
% 5.2 89.3970814000509\\
% };
% \addplot[forget plot, color=white!15!black] table[row sep=crcr] {%
% 0 0\\
% 6 0\\
% };
% %\addlegendentry{2 panels}

\addplot[ybar, xshift=+0.1cm, opacity=1, bar width=0.2, fill=mycolor2, draw=black, area legend] table[row sep=crcr] {%
0.8 0\\
2 0\\
3 0\\
4 0\\
5.2 0\\
};
\addplot[forget plot, color=white!15!black] table[row sep=crcr] {%
0 0\\
6 0\\
};
%\addlegendentry{3 panels}

\addplot[ybar, xshift=+0.1cm, opacity=1, bar width=0.2, fill=mycolor3, draw=black, area legend] table[row sep=crcr] {%
0.8 0\\
2 \\
3 0\\
4 0\\
5.2 79.9111533822713\\
};
\addplot[forget plot, color=white!15!black] table[row sep=crcr] {%
0 0\\
6 0\\
};
%\addlegendentry{4 panels}

\end{axis}

\pgfplotsset{ticks=none}
\begin{axis}[%
ybar,
width=0.951\fwidth,
height=\fheight,
at={(0\fwidth,0\fheight)},
scale only axis,
xtick=data,
%xmax=3000,
%xlabel style={font=\scriptsize\color{white!15!black}},
%xlabel={Distance [m]},
%xmajorgrids,
ymin=0,
ymax=100,
%ylabel style={font=\scriptsize\color{white!15!black}},
%ylabel={End-to-end latency [ms]},
%ymajorgrids,
ylabel shift = -5 pt,
yticklabel shift = -2 pt,
legend style={font=\scriptsize,at={(0.5,0.99)},anchor=north,legend cell align=left,align=left,draw=white!15!black},
enlarge x limits=0.15,
hide y axis,
hide x axis,
legend columns=2,
]
\addplot [fill=white,postaction={pattern=dots}]
  table[row sep=crcr]{%
  1 0\\
  2 0\\
  3 0\\
  4 0\\
};
\addlegendentry{UMi}

\addplot [fill=white]
  table[row sep=crcr]{%
  1 0\\
  2 0\\
  3 0\\
  4 0\\
};
\addlegendentry{UMa}
\end{axis}

\end{tikzpicture}%
    \caption{Average PDCP latency for different configurations of the sectors, distances $d$ and number of users $N_{UE}$.}
    \label{fig:latency}
\end{figure}

\begin{figure*}
\centering
	\setlength\belowcaptionskip{-0.45cm}

\begin{subfigure}[t]{0.48\textwidth}
	\centering
	\setlength\belowcaptionskip{-.15cm}

	\setlength\fwidth{0.85\columnwidth}    
	\setlength\fheight{0.55\columnwidth}
\definecolor{mycolor1}{rgb}{0.00000,0.44700,0.74100}%
\definecolor{mycolor4}{rgb}{0.6784,0.8471,0.9020}%
\definecolor{mycolor2}{rgb}{0.85000,0.32500,0.09800}%
\definecolor{mycolor5}{rgb}{1.0000,0.6275,0.4784}%
\definecolor{mycolor3}{rgb}{0.92900,0.69400,0.12500}%
\definecolor{mycolor6}{rgb}{1.0000,0.9804,0.8039}%
\begin{tikzpicture}
\pgfplotsset{every tick label/.append style={font=\scriptsize}}

\begin{axis}[%
% width=0.951\fwidth,
% height=\fheight,
% at={(0\fwidth,0\fheight)},
% scale only axis,
% bar shift auto,
% xmin=0,
% xmax=6,
% xtick={0,1,2,3,4,5,6},
% xticklabels={{},{2 uePanel, UMi},{3 uePanel, UMi},{},{2 uePanel, UMa},{3 uePanel, UMa},{}},
% ymin=0,
% ymax=40,
% ylabel style={font=\color{white!15!black}},
% ylabel={PDCP latency [ms]},
% axis background/.style={fill=white},
% axis x line*=bottom,
% axis y line*=left,
% legend style={legend cell align=left, align=left, draw=white!15!black}
% ]
width=0.951\fwidth,
height=\fheight,
at={(0\fwidth,0\fheight)},
axis line style={-},
scale only axis,
axis y line*=right,
% axis x line=none,%axis on top,
% hide x axis,
bar shift auto,
xmin=0,
xmax=6,
ymin=20,
ymax=42,
ylabel style={font=\scriptsize\color{white!15!black}},
ylabel={Average PDCP latency [ms]},
% axis background/.style={fill=white},
xtick={0,0.8,2,4,5.2,6},
xticklabel style={text width=30, align=center, font=\scriptsize},
xticklabels={{},{2 panels UMi},{3 panels UMi},{2 panels UMa},{3 panels UMa},{}},
ylabel shift = -5 pt,
yticklabel shift = -2 pt,
% xlabel shift = -4 pt,
xticklabel shift = -1 pt,
% axis x line*=bottom,
% axis y line*=left,
xmajorgrids,
ymajorgrids,
legend style={legend cell align=left, align=left, draw=white!15!black}
]
\addplot[xshift=-.1cm, ybar, postaction={pattern=dots}, bar width=0.229, fill=mycolor2, draw=black, area legend] table[row sep=crcr] {%
0.8	0\\
2	0\\
3	0\\
4	39.3267148030508\\
5.2	35.6121148125121\\
};
\addplot[forget plot, color=white!15!black] table[row sep=crcr] {%
0	0\\
6	0\\
};
% \addlegendentry{3 panels}

\addplot[xshift=-.1cm, ybar, postaction={pattern=dots}, bar width=0.229, fill=mycolor3, draw=black, area legend] table[row sep=crcr] {%
0.8	0\\
2	0\\
3	0\\
4	33.1225084592977\\
5.2	0\\
};
\addplot[forget plot, color=white!15!black] table[row sep=crcr] {%
0	0\\
6	0\\
};
% \addlegendentry{4 panels}

\end{axis}

% \begin{axis}[%
% width=0.951\fwidth,
% height=\fheight,
% at={(0\fwidth,0\fheight)},
% scale only axis,
% bar shift auto,
% xmin=0,
% xmax=6,
% xtick={0,1,2,3,4,5,6},
% xticklabels={{},{2 uePanel, UMi},{3 uePanel, UMi},{},{2 uePanel, UMa},{3 uePanel, UMa},{}},
% ymin=90,
% ymax=96,
% ylabel style={font=\color{white!15!black}},
% ylabel={PDCP throughput},
% % axis background/.style={fill=white},
% axis x line*=bottom,
% axis y line*=left,
% legend style={legend cell align=left, align=left, draw=white!15!black}
% ]
\begin{axis}[%
width=0.951\fwidth,
height=\fheight,
at={(0\fwidth,0\fheight)},
scale only axis,
bar shift auto,
axis x line=none,%axis on top,
xtick={0,0.8,2,4,5.2,6},
xmin=0,
xmax=6,
ymin=90,
ymax=98,
xlabel style={font=\scriptsize\color{white!15!black}},
ylabel style={font=\scriptsize\color{white!15!black}},
ylabel = {Average UDP user throughput [Mbit/s]},
ylabel shift = -5 pt,
yticklabel shift = -2 pt,
% xlabel shift = -4 pt,
xticklabel shift = -1 pt,
axis x line*=none,
axis y line*=none,
legend style={font=\scriptsize,legend cell align=left,align=left,draw=white!15!black, at={(0.01,0.99)},anchor=north west},
]
\addplot[ybar, bar width=0.229, fill=mycolor2, draw=black, area legend] table[row sep=crcr] {%
0.8	94.8310223164927\\
2	95.1605211197372\\
3	0\\
4	90.9073688089458\\
5.2	92.7130546237671\\
};
\addplot[forget plot, color=white!15!black] table[row sep=crcr] {%
0	0\\
6	0\\
};
\addlegendentry{3 sectors}

\addplot[ybar, bar width=0.229, fill=mycolor3, draw=black, area legend] table[row sep=crcr] {%
0.8	95.4968099548586\\
2	96.9554172858225\\
3	0\\
4	92.5002541975188\\
5.2	94.0194331769579\\
};
\addplot[forget plot, color=white!15!black] table[row sep=crcr] {%
0	0\\
6	0\\
};
\addlegendentry{4 sectors}

\end{axis}

\begin{axis}[%
% width=0.951\fwidth,
% height=\fheight,
% at={(0\fwidth,0\fheight)},
% scale only axis,
% bar shift auto,
% xmin=0,
% xmax=6,
% xtick={0,1,2,3,4,5,6},
% xticklabels={{},{2 uePanel, UMi},{3 uePanel, UMi},{},{2 uePanel, UMa},{3 uePanel, UMa},{}},
% ymin=0,
% ymax=40,
% ylabel style={font=\color{white!15!black}},
% ylabel={PDCP latency [ms]},
% axis background/.style={fill=white},
% axis x line*=bottom,
% axis y line*=left,
% legend style={legend cell align=left, align=left, draw=white!15!black}
% ]
width=0.951\fwidth,
height=\fheight,
at={(0\fwidth,0\fheight)},
axis line style={-},
scale only axis,
axis y line=none,
axis x line=none,%axis on top,
% hide x axis,
bar shift auto,
xmin=0,
xmax=6,
xtick={0,0.8,2,4,5.2,6},
ymin=20,
ymax=42,
ylabel style={font=\scriptsize\color{white!15!black}},
% ylabel={Average PDCP latency [ms]},
% axis background/.style={fill=white},
ylabel shift = -5 pt,
yticklabel shift = -2 pt,
% xlabel shift = -4 pt,
xticklabel shift = -1 pt,
% axis x line*=bottom,
% axis y line*=left,
% xmajorgrids,
% ymajorgrids,
]
\addplot[xshift=-.1cm, ybar, postaction={pattern=dots}, bar width=0.229, fill=mycolor2, draw=black, area legend] table[row sep=crcr] {%
0.8	27.2767434135122\\
2	26.1803946066957\\
3	0\\
4	0\\
5.2	0\\
};
\addplot[forget plot, color=white!15!black] table[row sep=crcr] {%
0	0\\
6	0\\
};
% \addlegendentry{3 panels}

\addplot[xshift=-.1cm, ybar, postaction={pattern=dots}, bar width=0.229, fill=mycolor3, draw=black, area legend] table[row sep=crcr] {%
0.8	23.6720623193619\\
2	20.82942072575220\\
3	0\\
4	0\\
5.2	0\\
};
\addplot[forget plot, color=white!15!black] table[row sep=crcr] {%
0	0\\
6	0\\
};

\end{axis}

\begin{axis}[%
% width=0.951\fwidth,
% height=\fheight,
% at={(0\fwidth,0\fheight)},
% scale only axis,
% bar shift auto,
% xmin=0,
% xmax=6,
% xtick={0,1,2,3,4,5,6},
% xticklabels={{},{2 uePanel, UMi},{3 uePanel, UMi},{},{2 uePanel, UMa},{3 uePanel, UMa},{}},
% ymin=0,
% ymax=40,
% ylabel style={font=\color{white!15!black}},
% ylabel={PDCP latency [ms]},
% axis background/.style={fill=white},
% axis x line*=bottom,
% axis y line*=left,
% legend style={legend cell align=left, align=left, draw=white!15!black}
% ]
width=0.951\fwidth,
height=\fheight,
at={(0\fwidth,0\fheight)},
axis line style={-},
scale only axis,
axis y line=none,
axis x line=none,%axis on top,
% hide x axis,
bar shift auto,
xmin=0,
xmax=6,
xtick={0,0.8,2,4,5.2,6},
ymin=20,
ymax=42,
ylabel style={font=\scriptsize\color{white!15!black}},
% ylabel={Average PDCP latency [ms]},
% axis background/.style={fill=white},
ylabel shift = -5 pt,
yticklabel shift = -2 pt,
% xlabel shift = -4 pt,
xticklabel shift = -1 pt,
% axis x line*=bottom,
% axis y line*=left,
% xmajorgrids,
% ymajorgrids,
]
\addplot[xshift=.3cm, ybar, postaction={pattern=dots}, bar width=0.229, fill=mycolor3, draw=black, area legend] table[row sep=crcr] {%
0.8	0\\
2	0\\
3	0\\
4	0\\
5.2	25.8700282963078\\
};
\addplot[forget plot, color=white!15!black] table[row sep=crcr] {%
0	0\\
6	0\\
};
\end{axis}

\pgfplotsset{ticks=none}
\begin{axis}[%
ybar,
width=0.951\fwidth,
height=\fheight,
at={(0\fwidth,0\fheight)},
scale only axis,
xtick=data,
%xmax=3000,
%xlabel style={font=\scriptsize\color{white!15!black}},
%xlabel={Distance [m]},
%xmajorgrids,
ymin=0,
ymax=100,
%ylabel style={font=\scriptsize\color{white!15!black}},
%ylabel={End-to-end latency [ms]},
%ymajorgrids,
ylabel shift = -5 pt,
yticklabel shift = -2 pt,
legend style={font=\scriptsize,at={(0.5,0.99)},anchor=north,legend cell align=left,align=left,draw=white!15!black},
enlarge x limits=0.15,
hide y axis,
hide x axis,
legend columns=2,
]
\addplot [fill=white,postaction={pattern=crosshatch dots}]
  table[row sep=crcr]{%
  1 0\\
  2 0\\
  3 0\\
  4 0\\
};
\addlegendentry{Latency}

\addplot [fill=white]
  table[row sep=crcr]{%
  1 0\\
  2 0\\
  3 0\\
  4 0\\
};
\addlegendentry{Throughput}
\end{axis}

\end{tikzpicture}%
    \caption{Average user throughput and latency for different configurations of the UE panels and gNB sectors, for distance $d = 100$ m and $N_{UE} = 25$ users. The \gls{3gpp} channel is either UMi or UMa.}
    \label{fig:panelsThLat}
\end{subfigure}\hfill
\begin{subfigure}[t]{0.48\textwidth}
	\centering
		\setlength\belowcaptionskip{-.15cm}

	\setlength\fwidth{0.9\columnwidth}    
	\setlength\fheight{0.53\columnwidth}
    \input{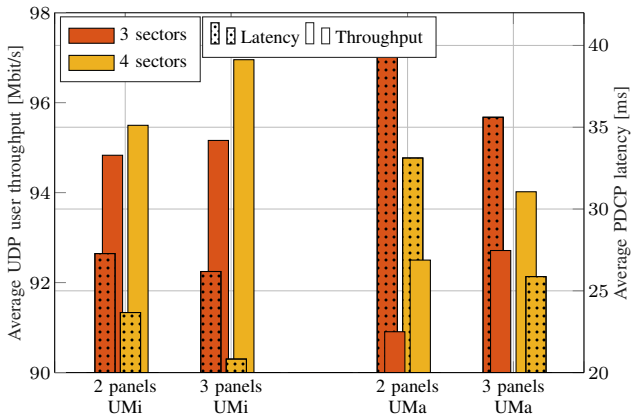}
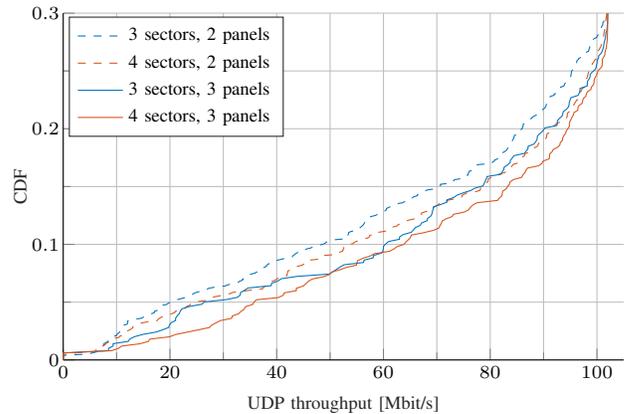
    \caption{CDF of the user UDP throughput for a configuration with 2 or 3 panels at the UE, different gNB sectors, UMa \gls{3gpp} channel configurations, and 25 UEs.}
    \label{fig:cdf}
\end{subfigure}
\caption{Comparison of the setup with 2 or 3 panels at the \gls{ue}.}
\label{fig:panels}
\end{figure*}

\vspace{-.15cm}
\subsection{Impact of the multi-sector deployment}
Figs.~\ref{fig:throughput} and~\ref{fig:latency} show the average UDP end-to-end throughput and \gls{ran} latency, respectively, for different numbers of sectors at the \glspl{gnb}, numbers of users $N_{UE}$ and distances $d$. The number of panels in each \gls{ue} is fixed to 2.

The first notable result, which holds for both UMa and UMi scenarios, is that the throughput increases when increasing the number of sectors in each \gls{gnb} from 3 to 4, while the average latency decreases. This is due to a combination of two factors. First, with more sectors it is possible to limit the angular coverage area of each sector, thus beams with a better shape and a higher gain are selected. Second, the interference decreases, since the usage of a multi-sector deployment limits the back and side lobes that generate undesired interference. The end result is an increase in the \gls{sinr}, which translates into higher throughput and lower latency, given that fewer retransmissions are needed and less buffering occurs.
Notice that, on average, the throughput gain is less remarkable than the latency reduction. This is due to the fact that the source rate is limited to $R_{UE} = 100$ Mbit/s, and most of the users experience an average good channel condition and can reach this throughput. The improvement is more relevant for the worst users, i.e., those who generally need a larger number of retransmissions, as we will show in the next paragraphs.
Finally, we highlight that increasing the number of sectors has a cost related to the \gls{gnb} hardware.

The second observation is that the UMi channel condition yields higher throughput and lower latency than UMa. The latter, indeed, generates a larger amount of interference across neighboring cells, thus decreasing the \gls{sinr}. The UMi scenario, instead, models a street canyon deployment, thus the inter-cell interference is much more limited.
%In fact, interference affect the performance of the transmission reducing the average throughput and increasing the number of RLC retransmission, thus increasing average user latency.  

\subsection{Impact of multiple panels at the \glspl{ue}}
We have shown in the previous result how the multi-sector \gls{gnb} deployment improves the link budget performance thanks to the possibility to better control the design of both desired and undesired beams. 
In this section, we consider different values for the number of panels at the \gls{ue}. As reported in Tab.~\ref{table_3gpp_antenna_settings}, the \gls{3gpp} suggests the use of 2 panels for each \gls{ue}. However, given the importance of handset design in 5G mmWave networks~\cite{huo2017cellular,huo2018cellular}, we evaluate the end-to-end performance by also configuring two different numbers of panels (i.e., 2 or 3) at the \gls{ue}. The results are reported in Fig.~\ref{fig:panelsThLat}.

Similar to the multi-sector deployment, the performance improves when installing 3 instead of 2 panels at the \glspl{ue}, for the throughput but more remarkably for the latency.
In particular, for the UMi configuration with 3 panels and 4 sectors, it is possible to nearly reach (on average) the maximum throughput $R_{UE}$ without increasing the latency, which has the smallest value with this configuration (i.e., 20.83~ms).

Furthermore, as mentioned in the previous section, the performance improvement can mostly be seen for the worst \glspl{ue}. Therefore, in Fig.~\ref{fig:cdf} we report the \gls{cdf} of the throughput for the configuration with 2 or 3 panels at the \gls{ue} and varying the number of \gls{gnb} sectors, for the UMa \gls{3gpp} channel configuration.
The plot shows that, indeed, almost 70\% of the users reach the saturation point (i.e., maximum achievable throughput).
Moreover, when comparing 2 or 3 panels at the \glspl{ue}, i.e., the dashed and the solid lines (of the same color), it can be seen that there is an improvement of up to 16~Mbit/s for the 10th percentile. In addition, the gain given by the larger number of panels is generally higher than that given by increasing the number of sectors.

However, even if the use of 3 panels at the \gls{ue} node results in an improvement of the performance, from a practical implementation point of view, the design of a \gls{ue} with these many panels must be studied carefully, since it may not be easy to physically place all the panels in the handset. Some preliminary designs and considerations are given in~\cite{huo2017cellular,huo2018cellular}.

% \begin{figure}
% 	\centering
% 	\setlength\fwidth{0.95\columnwidth}    
% 	\setlength\fheight{0.6\columnwidth}
%     \input{sectors-matlab/cdf_th_ue3pan.tex}
%     \caption{CDF of the user UDP throughput for a configuration with 3 panels at the UE, different gNB sectors, 25 UEs. \michelep{which distance, 3gpp?}}
%     \label{fig:cdf}
% \end{figure}

\subsection{Comparison with the isotropic antenna array}

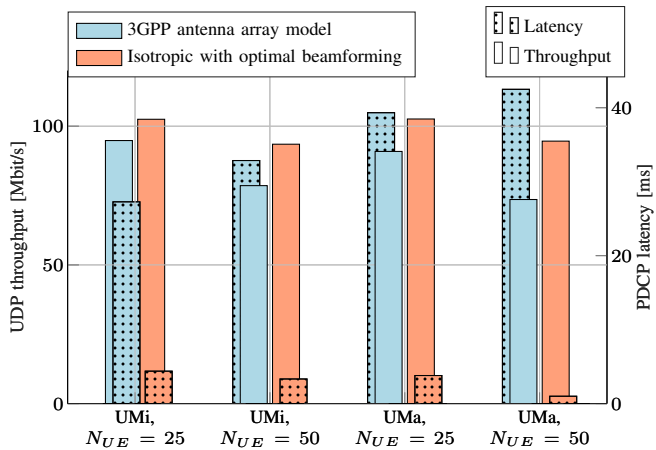
\begin{figure}
	\centering
	\setlength\fwidth{0.88\columnwidth}    
	\setlength\fheight{0.5\columnwidth}
\definecolor{mycolor1}{rgb}{0.00000,0.44700,0.74100}%
\definecolor{mycolor4}{rgb}{0.6784,0.8471,0.9020}%
\definecolor{mycolor2}{rgb}{0.85000,0.32500,0.09800}%
\definecolor{mycolor5}{rgb}{1.0000,0.6275,0.4784}%
\definecolor{mycolor3}{rgb}{0.92900,0.69400,0.12500}%
\definecolor{mycolor6}{rgb}{1.0000,0.9804,0.8039}%%
\begin{tikzpicture}
\pgfplotsset{every tick label/.append style={font=\scriptsize}}

\begin{axis}[%
width=0.92\fwidth,
height=\fheight,
at={(0\fwidth,0\fheight)},
scale only axis,
bar shift auto,
xmin=0.5,
xmax=4.5,
xtick={1,2,3,4},
xticklabel style={text width=50, align=center, font=\scriptsize},
xticklabels={{UMi, $N_{UE}=25$},{UMi, $N_{UE}=50$},{UMa, $N_{UE}=25$},{UMa, $N_{UE}=50$}},
ymin=0,
ymax=120,
ylabel style={font=\scriptsize\color{white!15!black}},
ylabel={UDP throughput [Mbit/s]},
axis background/.style={fill=white},
axis x line*=bottom,
axis y line*=left,
xmajorgrids,
ymajorgrids,
ylabel shift = -5 pt,
yticklabel shift = -2 pt,
% xlabel shift = -4 pt,
xticklabel shift = -1 pt,
legend style={font=\scriptsize,legend cell align=left,align=left,draw=white!15!black, at={(0.5,1.05)},anchor=south},
]
\addplot[ybar, bar width=0.2, xshift=0cm, fill=mycolor4, draw=black, area legend] table[row sep=crcr] {%
1	94.8310223164927\\
2	0\\
3	0\\
4 0\\
};
\addplot[forget plot, color=white!15!black] table[row sep=crcr] {%
0	0\\
5	0\\
};
%\addlegendentry{3GPP antenna array model}

\addplot[ybar, bar width=0.2, xshift=0cm, fill=mycolor5, draw=black, area legend] table[row sep=crcr] {%
1	102.5\\
2	93.51\\
3	102.6\\
4 94.62\\
};
\addplot[forget plot, color=white!15!black] table[row sep=crcr] {%
0	0\\
5	0\\
};
%\addlegendentry{Isotropic with optimal beamforming}

\end{axis}

\begin{axis}[%
width=0.92\fwidth,
height=\fheight,
at={(0\fwidth,0\fheight)},
scale only axis,
bar shift auto,
xmin=0.5,
xmax=4.5,
xtick={1,2,3,4},
ymin=0,
ymax=45,
ylabel shift = -5 pt,
yticklabel shift = -2 pt,
% xlabel shift = -4 pt,
xticklabel shift = -1 pt,
ylabel style={font=\scriptsize\color{white!15!black}},
ylabel={PDCP latency [ms]},
% axis background/.style={fill=white},
axis x line=none,
axis y line*=right,
legend style={legend cell align=left, align=left, draw=white!15!black}
]
\addplot[ybar, xshift=0.1cm, postaction={pattern=dots}, bar width=0.2, fill=mycolor4, draw=black, area legend] table[row sep=crcr] {%
1	27.2767434135122\\
2	0\\
3	0\\
4	0\\
};
\addplot[forget plot, color=white!15!black] table[row sep=crcr] {%
0	0\\
5	0\\
};

\addplot[ybar, xshift=0.1cm, postaction={pattern=dots}, bar width=0.2, fill=mycolor5, draw=black, area legend] table[row sep=crcr] {%
1	4.384\\
2	3.327\\
3	3.793\\
4	0.9933\\
};
\addplot[forget plot, color=white!15!black] table[row sep=crcr] {%
0	0\\
5	0\\
};

\end{axis}

\begin{axis}[%
width=0.92\fwidth,
height=\fheight,
at={(0\fwidth,0\fheight)},
scale only axis,
bar shift auto,
xmin=0.5,
xmax=4.5,
xtick={1,2,3,4},
ymin=0,
ymax=45,
ylabel shift = -5 pt,
yticklabel shift = -2 pt,
% xlabel shift = -4 pt,
xticklabel shift = -1 pt,
ylabel style={font=\scriptsize\color{white!15!black}},
ylabel={PDCP latency [ms]},
% axis background/.style={fill=white},
axis x line=none,
axis y line*=right,
legend style={legend cell align=left, align=left, draw=white!15!black}
]
\addplot[ybar, xshift=-0.1cm, postaction={pattern=dots}, bar width=0.2, fill=mycolor4, draw=black, area legend] table[row sep=crcr] {%
1 0\\
2 32.8604611528982\\
3 39.3267148030508\\
4 42.4870467349923\\
};
\addplot[forget plot, color=white!15!black] table[row sep=crcr] {%
0 0\\
5 0\\
};

\addplot[ybar, xshift=-0.1cm, postaction={pattern=dots}, bar width=0.2, fill=mycolor5, draw=black, area legend] table[row sep=crcr] {%
1 0\\
2 0\\
3 0\\
4 0\\
};
\addplot[forget plot, color=white!15!black] table[row sep=crcr] {%
0 0\\
5 0\\
};

\end{axis}

\begin{axis}[%
width=0.92\fwidth,
height=\fheight,
at={(0\fwidth,0\fheight)},
scale only axis,
bar shift auto,
xmin=0.5,
xmax=4.5,
xtick={1,2,3,4},
xticklabel style={text width=50, align=center, font=\scriptsize},
xticklabels={{UMi, $N_{UE}=25$},{UMi, $N_{UE}=50$},{UMa, $N_{UE}=25$},{UMa, $N_{UE}=50$}},
ymin=0,
ymax=120,
ylabel style={font=\scriptsize\color{white!15!black}},
ylabel={UDP throughput [Mbit/s]},
% axis background/.style={fill=white},
axis x line*=bottom,
axis y line*=left,
xmajorgrids,
ymajorgrids,
ylabel shift = -5 pt,
yticklabel shift = -2 pt,
% xlabel shift = -4 pt,
xticklabel shift = -1 pt,
legend style={font=\scriptsize,legend cell align=left,align=left,draw=white!15!black, at={(0,0.98)},anchor=south west},
]
\addplot[ybar, xshift=0cm,  bar width=0.2, fill=mycolor4, draw=black, area legend] table[row sep=crcr] {%
1 0\\
2 78.5763544422887\\
3 90.9073688089458\\
4 73.5720508595527\\
};
\addplot[forget plot, color=white!15!black] table[row sep=crcr] {%
0 0\\
5 0\\
};
\addlegendentry{3GPP antenna array model}

\addplot[ybar, xshift=0cm,  bar width=0.2, fill=mycolor5, draw=black, area legend] table[row sep=crcr] {%
1 0\\
2 0\\
3 0\\
4 0\\
};
\addplot[forget plot, color=white!15!black] table[row sep=crcr] {%
0 0\\
5 0\\
};
\addlegendentry{Isotropic with optimal beamforming}

\end{axis}

% \begin{axis}[%
% width=0.92\fwidth,
% height=\fheight,
% at={(0\fwidth,0\fheight)},
% scale only axis,
% bar shift auto,
% xmin=0.5,
% xmax=4.5,
% xtick={1,2,3,4},
% ymin=0,
% ymax=120,
% ylabel style={font=\scriptsize\color{white!15!black}},
% axis x line=none,
% axis y line=none,
% xmajorgrids,
% ymajorgrids,
% ylabel shift = -5 pt,
% yticklabel shift = -2 pt,
% % xlabel shift = -4 pt,
% xticklabel shift = -1 pt,
% ]
% \addplot[ybar, xshift=.2cm, bar width=0.2, fill=mycolor4, draw=black, area legend] table[row sep=crcr] {%
% 1	0\\
% 2	0\\
% 3	0\\
% 4	73.5720508595527\\
% };
% \addplot[forget plot, color=white!15!black] table[row sep=crcr] {%
% 0	0\\
% 5	0\\
% };
% % \addlegendentry{3GPP antenna array model}

% \addplot[ybar, xshift=.2cm, bar width=0.2, fill=mycolor5, draw=black, area legend] table[row sep=crcr] {%
% 1	0\\
% 2	0\\
% 3	0\\
% 4 94.62\\
% };
% \addplot[forget plot, color=white!15!black] table[row sep=crcr] {%
% 0	0\\
% 5	0\\
% };
% % \addlegendentry{Isotropic with optimal beamforming}

% \end{axis}

\pgfplotsset{ticks=none}
\begin{axis}[%
ybar,
width=0.951\fwidth,
height=\fheight,
at={(0\fwidth,0\fheight)},
scale only axis,
xtick=data,
%xmax=3000,
%xlabel style={font=\scriptsize\color{white!15!black}},
%xlabel={Distance [m]},
%xmajorgrids,
ymin=0,
ymax=120,
%ylabel style={font=\scriptsize\color{white!15!black}},
%ylabel={End-to-end latency [ms]},
%ymajorgrids,
ylabel shift = -5 pt,
yticklabel shift = -2 pt,
legend style={font=\scriptsize,at={(1,0.98)},anchor=south east,legend cell align=left,align=left,draw=white!15!black},
enlarge x limits=0.15,
hide y axis,
hide x axis,
]
\addplot [fill=white,postaction={pattern=dots}]
  table[row sep=crcr]{%
  1 0\\
  2 0\\
  3 0\\
  4 0\\
};
\addlegendentry{Latency}

\addplot [fill=white]
  table[row sep=crcr]{%
  1 0\\
  2 0\\
  3 0\\
  4 0\\
};
\addlegendentry{Throughput}

\end{axis}

\end{tikzpicture}%
    \caption{Average user throughput and latency for the \gls{3gpp} antenna model and the isotropic array with optimal beamforming described in~\cite{zhang2017ns3}, for distance $d =100$ m, different numbers of users $N_{UE}$. The \gls{3gpp} channel is UMi or UMa, with 3 sectors and 2 panels for the \gls{3gpp} antenna array model.}
    \label{fig:iso}
\end{figure}

In the plot of Fig.~\ref{fig:iso} we compare the average user throughput and latency for the scenario with a multi-sector, multi-panel \gls{3gpp} configuration against that with a single sector isotropic antenna elements. In the first, the beamforming vectors are computed as described in Sec.~\ref{sec:bf}, while the latter uses the optimal beamformers given by the eigenvector of the largest eigenvalue of the channel matrix~\cite{zhang2017ns3,love2003equal}.
This last approach, which we consider as baseline, represents the default setting for antenna radiation and beamforming in the ns-3 mmWave module. 

This comparison highlights the gap in the performance between the two antenna array models considered.
The throughput is slightly lower (and, conversely, latency is higher) in the \gls{3gpp} antenna array model for both UMi and UMa scenarios.
This is due to the use of the optimal beamforming vectors together with the isotropic antenna model which result in a configuration that identifies an upper bound in the performance.
Indeed, even if the interference power irradiated in the side lobes of the isotropic antenna array model is bigger than in the \gls{3gpp} model, this cannot balance the advantage of the optimal beamforming vectors.
A detailed evaluation on the effects of the interference power irradiated can be found in~\cite{rebato18,rebato2016understanding}.

\vspace{-.3cm}
\section{Conclusions}
\label{sec:concl}
In this paper, we improved the channel model abstraction of the mmWave module for ns-3, by introducing the support of a more realistic antenna array model, compliant with \gls{3gpp} \gls{nr} requirements, which adopts multiple antenna arrays at the base stations and mobile handsets.
We then evaluated the end-to-end performance of a mmWave cellular network by varying channel and antenna array configurations.
Our results show that by increasing the number of sectors it is possible to improve the \gls{sinr}, thus increasing user throughput and at the same time decreasing latency.
Moreover, we highlight that the configuration with a single sector of isotropic antenna elements and optimal beamforming vectors has results which are slightly better than the configuration with \gls{3gpp} specifications in the scenario we consider, i.e., with a limited number of interfering sources.

Therefore, as a future extension of this work, we will further evaluate the end-to-end performance of networks with different multi-sector and multi-panel configurations, to clearly outline the trade-offs related to the antenna array configurations and their modeling.
For example, we will analyze larger simulation scenarios, i.e., with more \glspl{gnb} and \glspl{ue} deployed and consequently a higher interference, other antenna array factor components, such as the spacing of the elements and the amplitude and the phase vectors of each antenna element, and network configurations.

\vspace{-.3cm}
\bibliographystyle{IEEEtran}
\bibliography{bibl}

\end{document}